\documentclass[aps,twocolumn,showpacs,groupedaddress,nofootinbib]{revtex4-1}
\usepackage{graphicx}
\usepackage{amsmath}
\usepackage{amssymb}
\usepackage{color}

\newcommand{\p}{\textbf{p}}

\begin{document}

\title{Turbulent thermalization of weakly coupled non-abelian plasmas}
\author{S\"{o}ren Schlichting}
\affiliation{\vspace{0.2cm}
Institut f{\"u}r Theoretische Physik,Universit{\"a}t Heidelberg, Philosophenweg 16, 69120 Heidelberg
}

\begin{abstract}
We study the dynamics of weakly coupled non-abelian plasmas within the frameworks of classical-statistical lattice gauge-theory and kinetic theory.  We focus on a class of systems which are highly occupied, isotropic at all times and initially characterized by a single momentum scale. These represent an idealized version of the situation in relativistic heavy ion-collisions in the color-glass condensate picture, where on a time scale $1/Q_s$ after the collision of heavy nuclei a longitudinally expanding plasma characterized by the saturation scale $Q_s$ is formed. Our results indicate that the system evolves according to a turbulent Kolmogorov cascade in the classical regime.  Taking this into account, the kinetic description is able to reproduce characteristic features of the evolution correctly.
\end{abstract}

\maketitle

\section{Introduction}
Understanding the mechanisms underlying the fast thermalization observed in relativistic heavy-ion collisions \cite{THERM} provides one of the biggest theoretical challenges in our current understanding of the experiments performed at RHIC and the LHC \cite{EMMI}. Though there has been significant progress in the development of ab-initio calculations \cite{ABINITIO}, these have not yet been able to address the problems of thermalization and isotropization \cite{Glasma}. On the other hand several authors have studied thermalization of non-abelian plasmas in a more general context at strong \cite{strong} and weak coupling \cite{BMSS,AK1,Blaizot,AK2}. We follow the same approach and study the problem of thermalization from a more general point of view, while focussing on a class of systems which share important features with the non-equilibrium stage of relativistic heavy ion collisions. We will work at weak coupling $\alpha\ll1$ and consider a class of homogenous, isotropic sysyems which initially have a parametrically large occupation
\begin{eqnarray}
\label{eq:initial}
f(\p)\sim \alpha^{-c}\;\text{for}\;|\p|<Q\;, \quad f(|\p|>Q)\ll1,
\end{eqnarray}
where $0<c<1$, and are characterized by a single dimensionful scale $Q$. As discussed in Ref. \cite{Blaizot} this mimics the behavior in heavy-ion collisions at times $\tau\sim Q^{-1}$ after the collision of heavy nuclei. The main difference is the longitudinal expansion of the system, which we will neglect in the following. This greatly simplifies the problem, as the system can be considered isotropic at all times, whereas in the longitudinally expanding case the system remains anisotropic on large time scales \cite{Blaizot,AK2}. Thermalization of such a system has been studied in a kinetic theory framework in Refs. \cite{AK1,Blaizot} and first results from classical-statistical lattice simulations have been presented in Ref. \cite{prev}. While the authors of Refs. \cite{AK1} and \cite{Blaizot} agree on the way thermalization proceeds in these type of overoccupied systems, numerical simulations have raised the question whether alternatively thermalization may proceed as a turbulent cascade \cite{prev,BergesTurb}. The latter are characterized by non-thermal fixed points of classical field theories, which can be associated to stationary transport of conserved quantities \cite{TurbBook,TurbRev}. The occurrence of these solutions is well known for a variety of systems including early universe cosmology \cite{cosmo} and cold atomic gas systems \cite{qgas,qgas1,qgas2}, and turbulent behavior has also been discussed in the context of non-abelian gauge theories \cite{prev,BergesTurb,MooreTurb,gaugeTurb,gaugeTurb2}. In this paper we will focus on the dynamical evolution of the system and refer to the literature for analytic studies of turbulence \cite{TurbRev,TurbBook,prev,BergesTurb,MooreTurb,gaugeTurb,gaugeTurb2}. Our strategy is to combine kinetic theory and classical statistical lattice simulations in order to provide a simple understanding based on parametric estimates while having control over non-perturbative and non-equilibrium effects. We start with a kinetic theory analysis in Sec. \ref{sec:kin}, where we briefly review the discussion in Refs. \cite{AK1,Blaizot} and extend it to account for the presence of a non-thermal fixed point. In Sec. \ref{sec:lat} we present results from classical-statistical lattice simulations, which indicate the occurrence of a turblent cascade with exponent $\kappa\simeq4/3$ at late times, as observed in Ref. \cite{BergesTurb}. We then summarize our results and conclude with Sec. \ref{sec:conc}.
\section{Evolution in Kinetic theory}
\label{sec:kin}
The kinetic evolution of systems characterized with an initial distribution as in Eq. (\ref{eq:initial}) has been studied in Refs. \cite{AK1,Blaizot} and we will adopt the notation of Ref. \cite{Blaizot} to analyze the evolution of the system. To describe the evolution, one has to take into account the effects of elastic and inelastic scattering, as well as the interaction of hard and soft excitations. We start with a discussion of elastic scattering, which in this case is dominated by scattering of hard particles with small momentum transfer \cite{AK1,Blaizot}. In the small angle approximation, where the average momentum of hard excitations is much bigger than the average exchanged momentum, the process appears as a 'random walk' in momentum space, controlled by the momentum diffusion parameter (c.f. Ref. \cite{AK1,Blaizot})
\begin{eqnarray}
\label{eq:qhatel}
\hat{q}_{\text{el}}\sim\alpha^2\int\frac{d^3p}{(2\pi)^3}f(p)[1+f(p)] \;.
\end{eqnarray}
Similarly one can estimate the effects of inelastic processes and interactions of hard particles with soft gauge field modes below the Debye scale. The important observation is that for overoccupied, isotropic systems all processes are parametrically as efficient as elastic scattering \cite{AK1,Blaizot} such that
\begin{eqnarray}
\hat{q}\sim\hat{q}_{el}\sim\hat{q}_{inel}\sim\hat{q}_{soft}\;. 
\end{eqnarray}
This greatly simplifies the discussion of the kinetic evolution of the system, since we do not have to distinguish different regimes where either of the processes dominate. From Eq. (\ref{eq:qhatel}) we can also obtain the rate at which a particle with momentum $\p$ exhibits a momentum transfer of the same order. This rate is parametrically given by
\begin{eqnarray}
\label{eq:scatrates}
\Gamma_{el}(\p)\sim\frac{\hat{q}_{el}}{\p^2}\;, 
\end{eqnarray}
such that soft particles experience large angle scatterings at a higher rate as compared to hard particles. Given the enhanced scattering rates of soft excitations, we expect that the soft tail of the distribution can always 'equilibrate' to a distribution which is dictated by the dynamics of the hard modes. We will assume in the following that this distribution is described by a power law, where
\begin{eqnarray}
\label{eq:IsoInit}
f(\p)\sim\alpha^{-c} \left(\frac{\Lambda_s}{\omega_\p}\right)^{\kappa} \quad \text{for} \quad p<\Lambda
\end{eqnarray}
up to a momentum cutoff $\Lambda$ above which occupancies are negligible. Here the exponent $c$ characterizes parametrically the occupation of the system, and we will only consider the case $0<c<1$ of initially over-occupied systems.  The exponent $\kappa$ characterizes the shape of the distribution, while $\kappa=1$ corresponds to a thermal shape, the values $\kappa=3/2$ (c.f. Ref. \cite{prev}) and $\kappa=4/3,5/3$ (c.f. Ref. \cite{BergesTurb}) are known to appear for systems which exhibit Kolmogorov wave turbulence.%
\footnote{
The appearance of the exponent $\kappa=3/2$  can be explained with the existance of a classical background field \cite{prev,TurbRev}. This aspect is discussed in more detail in Ref. \cite{prev} in the context of the recent debate on condensate formation out-of equilibrium \cite{Blaizot,cond}.
}
The scales $\Lambda$ and $\Lambda_s$ in Eq. (\ref{eq:IsoInit}) both depend on time and determine the evolution of the system. At initial time $t_0\sim Q^{-1}$ the two scales coincide and one finds $\Lambda\sim\Lambda_s\sim Q$. In order to extract the time evolution of the scales $\Lambda$ and $\Lambda_s$ we first note that for $\kappa<3/2$ the momentum diffusion parameter is parametrically given by
\begin{eqnarray}
\label{eq:qhatk}
\hat{q}\sim\alpha^{2-2c}\left(\frac{\Lambda_s}{\Lambda}\right)^{2\kappa}\Lambda^3 \;.
\end{eqnarray}
In contrast if one considers $\kappa>3/2$ the integral in Eq. (\ref{eq:qhatel}) is dominated by infrared contributions indicating a break down of the kinetic description. We will therefore limit our discussion to the case $\kappa<3/2$, where we expect a kinetic description to apply. As we will see shortly in Sec. \ref{sec:lat}, this situation is also realized in our lattice simulations. From Eq. (\ref{eq:scatrates}) we can infer the time scale $t_{\text{start}}$ on which the momentum of initially hard excitations changes appreciably as
\begin{eqnarray}
t_{\text{start}}\sim Q^{-1} \alpha^{2c-2}\;. 
\end{eqnarray}
Before the time $t_{\text{start}}$ the system will develop its soft tail, which then moves outwards to higher momenta until at times $t\sim t_{\text{start}}$ changes in the distribution of hard excitations start to take place. In the regime $t\gtrsim t_{\text{start}}$ one can find a self-consistent scaling solution by requiring the evolution of the hard scale to be governed by momentum diffusion such that (c.f. Ref. \cite{AK1,Blaizot})
\begin{eqnarray}
\frac{d}{dt}\Lambda^2\sim\hat{q} \sim\alpha^{2-2c} \left(\frac{\Lambda_s}{\Lambda}\right)^{2\kappa}\Lambda^3\;.
\end{eqnarray}
We note that for non-expanding systems the ratio of soft and hard scale $(\Lambda_s/\Lambda)^{\kappa}$ is fixed by energy conservation
\begin{eqnarray}
e\sim\alpha^{-c}\left(\frac{\Lambda_s}{\Lambda}\right)^{\kappa}\Lambda^4\sim\alpha^{-c}Q^4 \;,
\end{eqnarray}
such that the relation
\begin{eqnarray}
\left(\frac{\Lambda_s}{\Lambda}\right)^{\kappa}\sim \left(\frac{Q}{\Lambda} \right)^4 \;,
\end{eqnarray}
holds at all times of the evolution. Combining the above conditions then yields the evolution of the scales $\Lambda$ and $\Lambda_s$ for times $t\gtrsim t_{\text{start}}$ to be
\begin{eqnarray}
\Lambda\sim Q \left(\frac{t}{t_\text{start}}\right)^{1/7}\;, \quad \Lambda_s\sim Q \left(\frac{t}{t_\text{start}}\right)^{(1-4/\kappa)/7}\;,
\end{eqnarray}
which agrees with the results obtained in Refs. \cite{AK1,Blaizot} for the case $\kappa=1$ of a (quasi-)thermal distribution. We note that the evolution of the hard scale $\Lambda$ as well as the evolution of the occupancies of hard modes $(\Lambda_s/\Lambda)^\kappa$  turn out to be independent of the exponent $\kappa$. This behavior is a direct consequence of energy conservation.\\
\\
The evolution of the scales $\Lambda$ and $\Lambda_s$ proceeds in this way until at some points the occupancies of hard modes become $\mathcal{O}(1)$, where quantum effects become important. These will have the effect of driving the system towards its unique thermal fixed point, and hence complete the thermalization process. We therefore expect thermalization of the system to occur as two step process, where the first regime is characterized by the above scaling solutions. The timescale $t_{\text{change}}$ for entering the second regime, where the 'rest-thermalization' takes place is parametrically given by 
\begin{eqnarray}
t_{\text{change}}\sim Q^{-1}\alpha^{-2+c/4} 
\end{eqnarray}
which is independent of the value of the exponent $\kappa$. However for $\kappa\neq1$ even at times $t\sim t_{\text{change}}$ the system still deviates significantly from its equilbirium distribution and one has to take into account also the time for the 'rest-thermalization' to occur. If the latter is assumed to be parametrically faster than our previous estimate one obtains $t_{\text{eq}}\sim t_{\text{change}}$ as the final result. In this situation one recovers the estimate of Refs. \cite{AK1,Blaizot} for the overall thermalization time.
\section{Evolution in classical-statistical lattice gauge-theory}
\label{sec:lat}
In this section we present the results from classical statistical lattice simulations of $SU(2)$ Yang-Mills theory. The primary goals are to analyze whether the above scaling solutions are realized in a genuine non-equilibrium approach and to determine the scaling exponent $\kappa$. \\
\\
When comparing the lattice results to the kinetic theory evolution, we have to overcome the challenge of identifying the quasi-particle excitations discussed in the kinetic theory framework. This is not unambiguous since the notion of quasi-particles is a perturbative concept and there is no general correspondence to  two-point correlation functions. However one can typically exploit the perturbative behavior of the two-point correlation functions to establish a quasi particle picture, in situations where this is applicable. With regard to non-abelian gauge-theories the problem becomes even more severe, due to the problems associated with gauge-invariance. Typically the particle number defined from two-point correlation functions will not be a gauge invariant quantity and it is therefore important to consider a gauge which has a clear interpretation in terms of physical degrees of freedom. In practice the choice of gauge is limited by the necessity to use temporal axial gauge ($A_t=0$) in the classical-statistical lattice simulations. However the gauge fixing $A_t=0$ is incomplete such that one has the residual freedom to perform time-independent gauge transformations. Fixing the residual gauge by an additional constraint eliminates gauge-fluctuations and can be used to define quasi-particle observables. We follow Ref. \cite{prev} and address this problem by implementing the Coulomb type constraint $\triangledown A=0$, whenever we are interested in gauge-dependent observables. We then study the evolution of the electric and magnetic two-point correlation functions
\begin{eqnarray}
\label{eq:corrfunc}
\langle BB \rangle(t,\textbf{p})&=&\int_{\textbf{x-y}}\frac{\langle B_i^a(t,\textbf{x}) B_i^a(t,\textbf{y})\rangle}{(N_c^2-1)(d-1)}e^{-i\textbf{p(x-y)}}\\
\label{eq:corrfunc1}
\langle EE \rangle(t,\textbf{p})&=&\int_{\textbf{x-y}}\frac{\langle E_i^a(t,\textbf{x}) E_i^a(t,\textbf{y})\rangle}{(N_c^2-1)(d-1)}e^{-i\textbf{p(x-y)}}
\end{eqnarray}
here written in terms of continuum variables, where $N_c$ is the number of colors, $d=3$ is the number of spatial dimensions and $\langle . \rangle$ denotes classical-statistical averaging.\footnote{The normalization factor $(N_c^2-1)(d-1)$ corresponds to the number of massless physical gluons. This normalization implicitly assumes that the unphysical degrees of freedom are removed by the gauge-fixing procedure.} The classical-statistical lattice implementation is describe in detail in Ref. \cite{simulations} and we use the so-called 'Los Alamos' gauge-fixing procedure \cite{gaugeFixing}. The initial conditions in our simulations are chosen to mimic the quasi-particle picture in Eq. (\ref{eq:initial}) and described in more detail in the appendix. The simulations are performed on $N=64,96,128$ hypercubic lattices with three different values of the lattice spacing $Qa=0.33,0.66,1.0$ to gain systematic control over lattice spacing and finite volume effects. If not stated otherwise results are presented for $N=96$ and $Qa=0.66$ lattices.
\begin{figure}[t]
\includegraphics[width=0.55\textwidth]{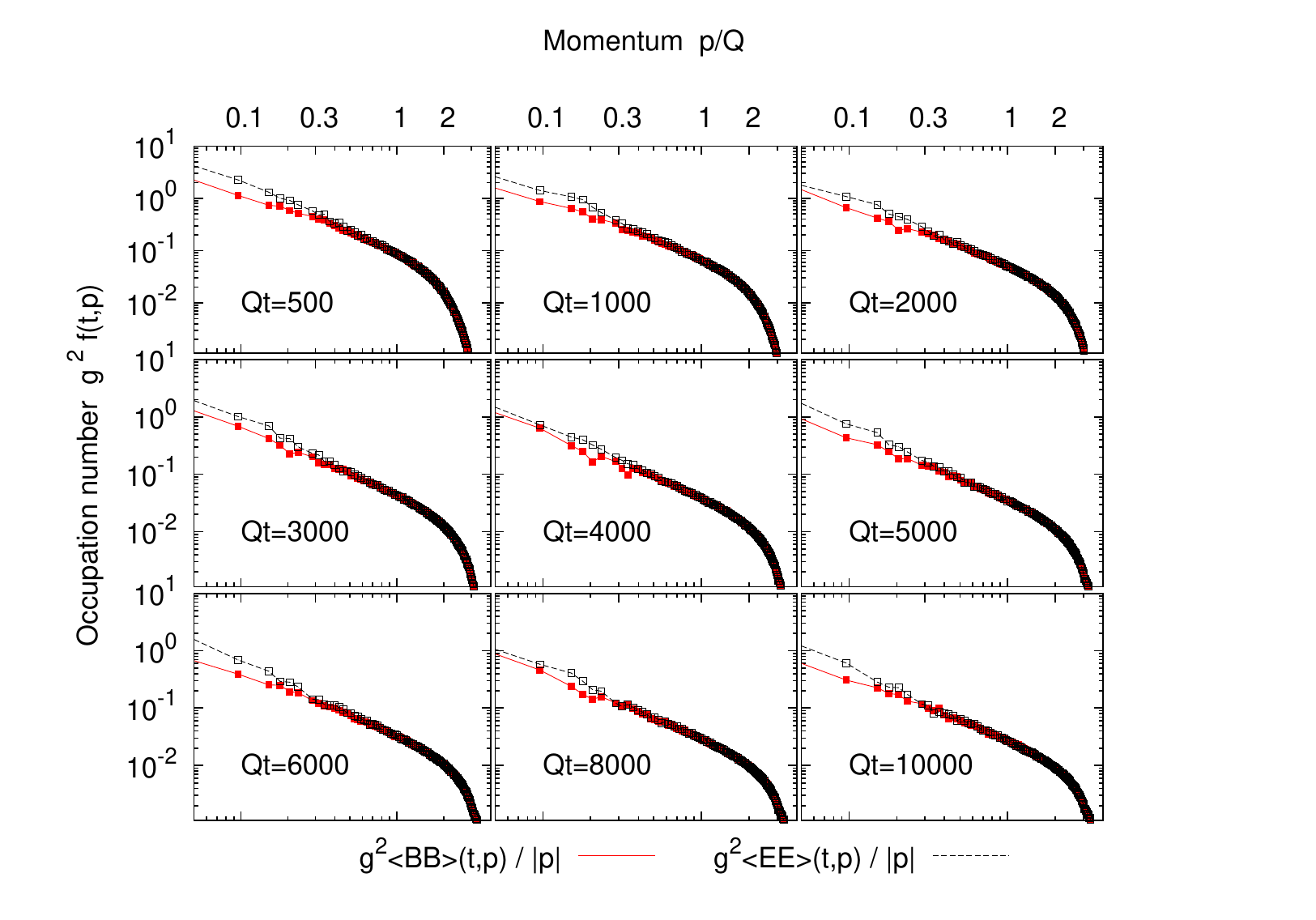}
\caption{\label{fig:Spectra} (color online) Spectrum of excitations obtained from the correlation functions, $\langle EE \rangle/|\p|$ and $\langle BB \rangle/|\p|$ at different times of the evolution. One observes a clear power-law dependence for the  correlation functions $\langle EE \rangle$ and $\langle BB \rangle$, which extends over approximately one decade. As time proceeds the tail of the power law propagates towards the ultraviolet, while the amplitude of the distribution decreases.}
\end{figure}  
\subsubsection{Highly overoccupied systems \\ ($f_0(p)\sim\alpha^{-1}$)}
\begin{figure}[t]
\includegraphics[width=0.45\textwidth]{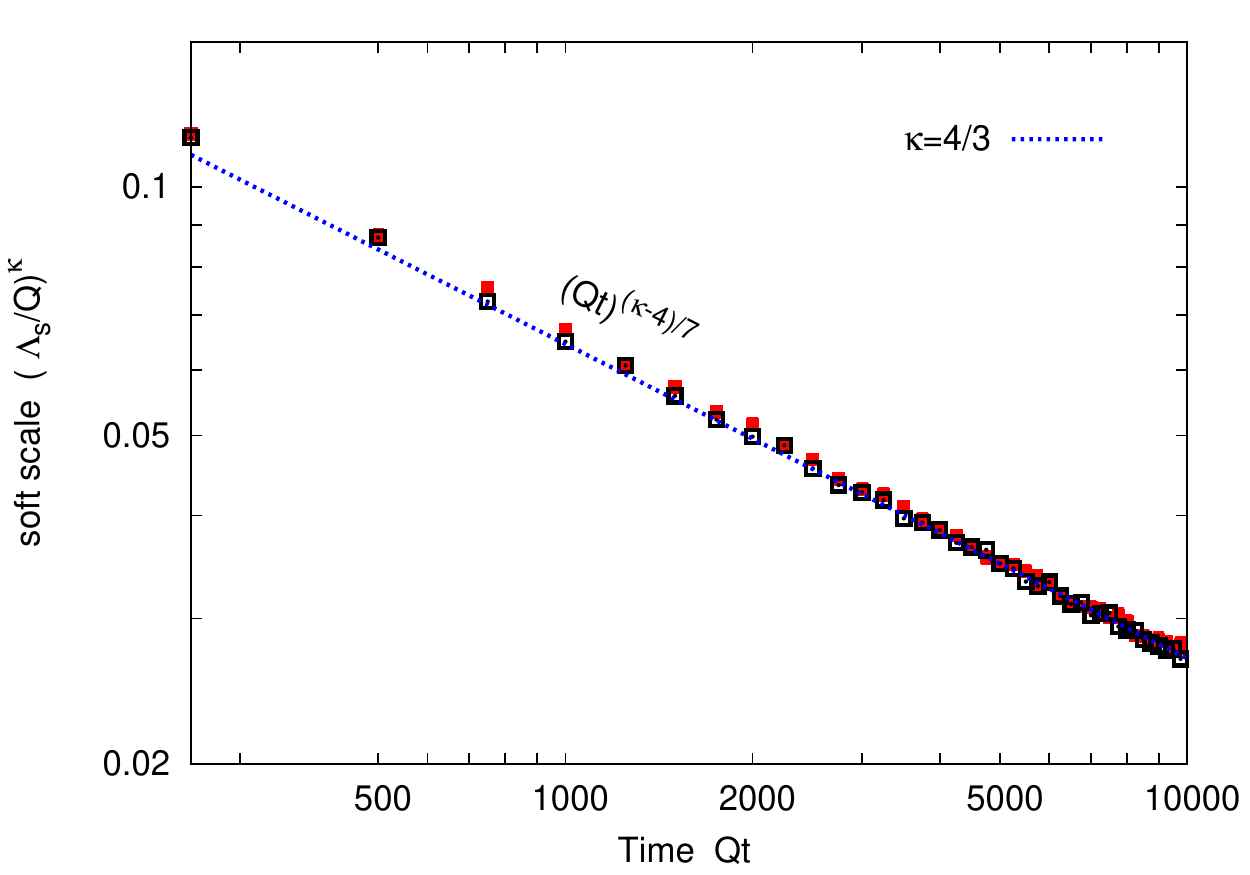}
\caption{\label{fig:LambdaSoft} (color online) Time evolution of the soft scale $(\Lambda_s/Q)^\kappa$ extracted from fits of the spectra at different times (red squares) and from the time evolution of the occupation number for modes with $p\simeq Q$ (black squares). The blue dashed line corresponds to a power-law behavior $(Qt)^{(\kappa-4)/7}$ as expected from the kinetic theory analysis, where we assumed $\kappa=4/3$ as a constant in time.}
\end{figure}

 We first study the time evolution of the correlation functions defined in Eq. (\ref{eq:corrfunc}) and (\ref{eq:corrfunc1}). The results are presented in Fig. \ref{fig:Spectra}, where we show the spectrum of the correlation functions $\langle EE \rangle(t,\p)/|\p|$ and $\langle BB \rangle(t,\p)/|\p|$ at different times of the evolution. The factor  $1/|\p|$ is chosen to produce dimensionless quantities, which can be interpreted as an occupation numbers. From Fig. \ref{fig:Spectra} one observes the build up of a power law distribution, which subsequently decreases in amplitude while slowly moving out towards higher momenta. We observe that towards later times the electric and magnetic correlation functions $\langle EE \rangle/|\p|$ and $\langle BB \rangle/|\p|$ show a consistent scaling behavior over a wide range of momenta. In order to establish a more direct comparison with the kinetic theory discussion we define the effective occupation number from the electric field correlation function as
\begin{eqnarray}
\label{eq:defOcc}
f(t,\p)=\frac{\langle E E \rangle(t,\p)}{|\p|}\;, 
\end{eqnarray}
which we will use in the following to further analyze the evolution.\\ 
\\
\begin{figure}[t]
\includegraphics[width=0.45\textwidth]{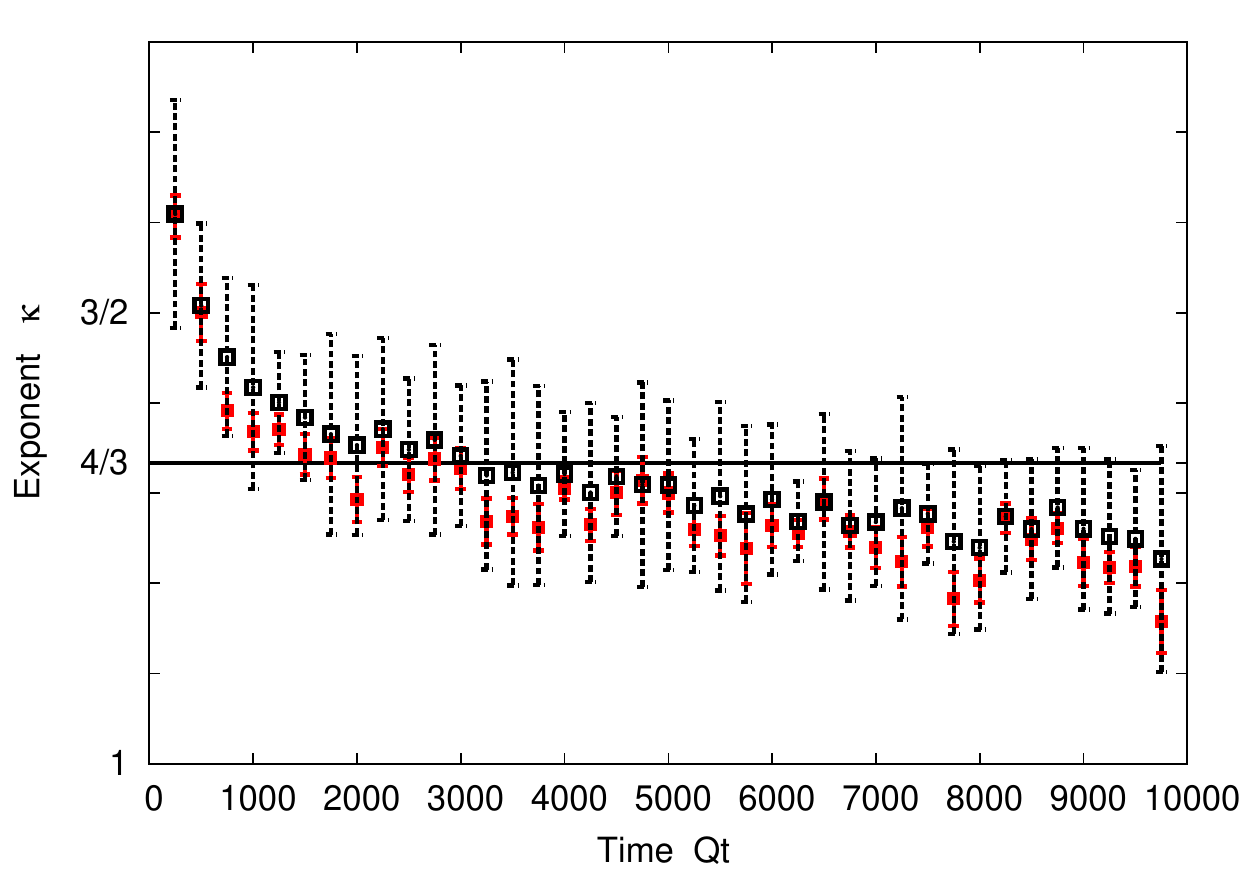}
\caption{\label{fig:Exponent} (color online) Scaling exponent $\kappa$ extracted from the spectrum of the correlation function $f(t,\p)\sim\langle EE\rangle/|\p|$ at different times of the evolution. The different symbols correspond to the two different extraction methods. The red filled squares correspond to a global fit in the momentum range $p=0.3-1.0~Q$, whereas the black empty squares correspond to average of local fits as described in the text. Over a large time-scale the exponent is consistent with the Kolmogorov turbulence exponent $\kappa=4/3$ \cite{BergesTurb}.}
\end{figure}
While the power law dependence of the distribution function $f(t,\p)$ can already be observed from Fig. \ref{fig:Spectra} we are interested also in the exponent $\kappa$ of the power law. This is important to distinguish the cases where the observed spectrum corresponds to a quasi-thermal evolution ($\kappa=1$) or a turbulent Kolmogorov cascade ($\kappa=4/3,5/3;~3/2$). To extract the exponent from the data we perform a series of least-square fits at each time slice, with a distribution function of the form
\begin{eqnarray}
\label{eq:fScaling}
f(t,\p)=\alpha^{-c} \left(\frac{\Lambda_s(t)}{|\p|}\right)^{\kappa(t)}\;. 
\end{eqnarray}
This procedure yields the two parameters $\kappa(t)$ and $(\Lambda_s/Q)^\kappa(t)$ which are shown in Fig. \ref{fig:Exponent} and \ref{fig:LambdaSoft} respectively. To estimate the error of this procedure we perform the analysis in two different ways: In the first case we simply consider the result of a global fit along with its error in the momentum range $p/Q=0.3-1.0$, where scaling is observed. The results of this procedure are shown as red squares in Fig. \ref{fig:Exponent}. In the second case we divide the data in momentum bins and perform a separate fit for each bin. We then extract the average exponent and its error in the scaling region. The results correspond to the black points in Fig. \ref{fig:Exponent}. One observes that the results for $\kappa$ of the two procedures agree, while the second method provides a more reliable estimate of the error. We find that at early times the spectrum features a power law exponent of $\kappa\simeq3/2$, which is consistent with previous observations \cite{prev}. At later times we observe a hardening of the spectrum where the data is in favor of $\kappa\simeq4/3$ over a large time-scale. The appearance of the exponent $\kappa=3/2$, may be attributed to a transient condensation phenomenon, which does not persist on parametrically large time scales \cite{prev,cond}.  The thermal value $\kappa=1$ is clearly ruled out by the data over the entire simulation time.\\
\\
As discussed in Sec. \ref{sec:kin} in the kinetic theory framework, the appearance of a non-thermal exponent ($\kappa\neq1$) has an immediate impact on the evolution of the soft scale $\Lambda_s$, which is expected to display a slower evolution for larger values of $\kappa$. To investigate whether this is supported by the lattice data, we extract the time evolution of the soft scale $(\Lambda_s/Q)^\kappa(t)$ by two different procedures. The results are on display in Fig. \ref{fig:LambdaSoft}. In the first case we use the results of the previous fit procedure to directly obtain the quantity $(\Lambda_s/Q)^{\kappa}$ as a function of time. This corresponds to red squares in Fig. \ref{fig:LambdaSoft}. In the second case we investigate the evolution of modes with $|\p|\simeq Q$, where $f(t,|\p|\simeq Q)\sim\alpha^{-c}(\Lambda_s/Q)^\kappa$ according to Eq. (\ref{eq:fScaling}). This is shown by black squares in Fig. \ref{fig:LambdaSoft}. The results from both methods agree and one observes a clear power law dependence. The blue dashed line corresponds to the kinetic theory estimate (c.f. Sec. \ref{sec:kin})
\begin{eqnarray}
\label{eq:softScaling}
\left(\frac{\Lambda_s(t)}{Q}\right)^{\kappa}\sim(Qt)^{(\kappa-4)/7} 
\end{eqnarray}
where we assumed $\kappa=4/3$ to be independent of time. One observes good overall agreement with the data. We also performed fits to the extract the scaling exponent from the time evolution of $(\Lambda_s/Q)^\kappa$. Using the same methods as introduced previously and assuming the relation (\ref{eq:softScaling}) this yields $\kappa\simeq1.3\pm0.1$, which is consistent with the values shown in Fig. \ref{fig:Exponent}. \\
\\
\begin{figure}[t]
\includegraphics[width=0.45\textwidth]{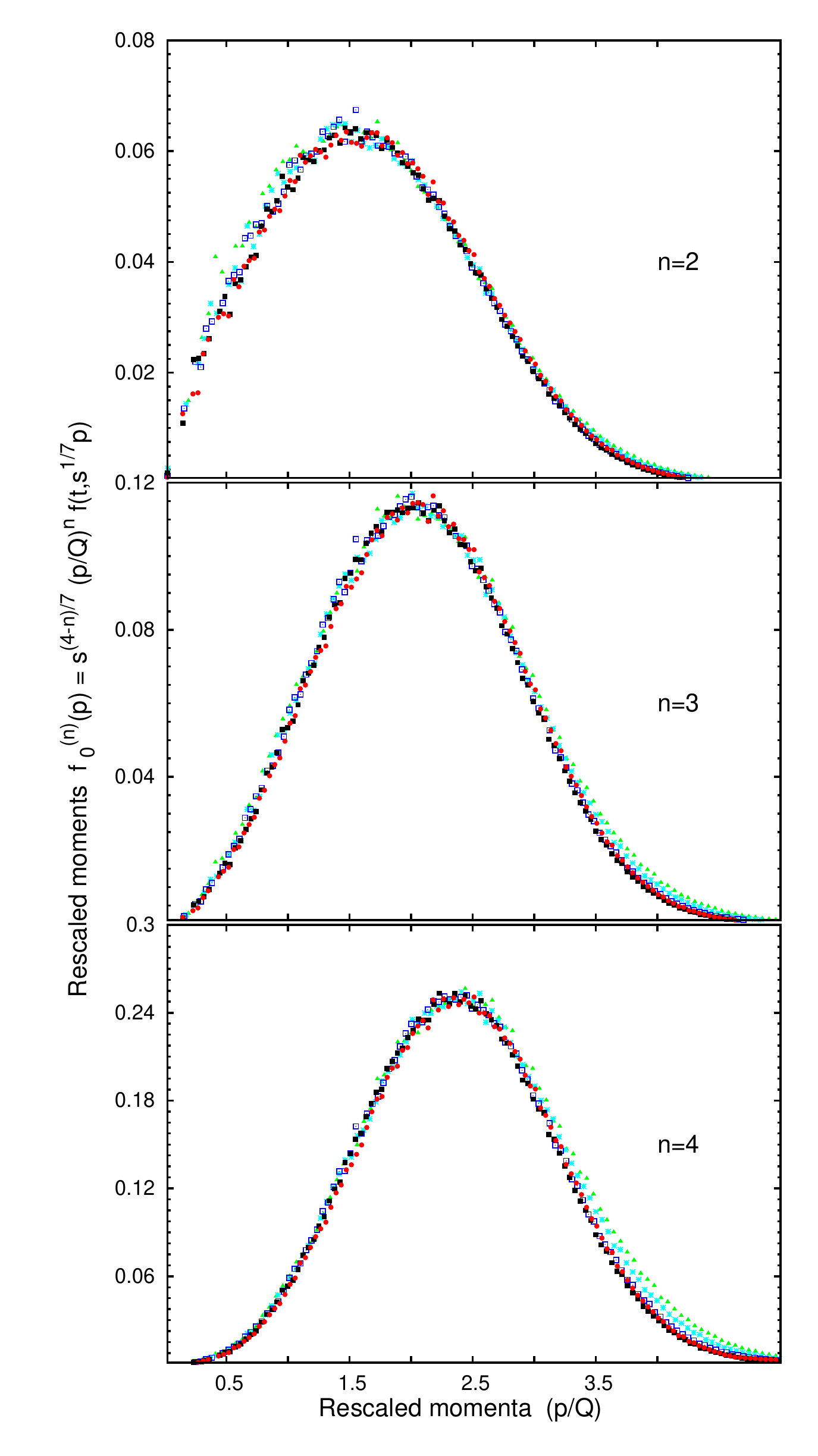}
\caption{\label{fig:Moments} (color online) Rescaled moments of the distribution function as a function of the rescaled momentum variable. The different colors and symbols correspond to different evolution times. The scale factor is $s=t/t_0$, where we chose $Qt_0=1000$ as normalization. The scaling exponents are sensitive to the evolution of the hard scale $\Lambda(t)$ and we used the values from the kinetic theory analysis in Sec. \ref{sec:kin}. The results are obtained on $N=96$ lattices with $Qa=0.33$.}
\end{figure} 
So far we have only analyzed the behavior of the spectrum in a momentum region where it shows a clear power law dependence. However one of the key features of the evolution is the propagation towards higher momenta, which ultimately leads to thermalization of the system. As discussed in Sec. \ref{sec:kin} the kinetic theory framework provides a strong prediction of how this evolution proceeds, which is insensitive also to the power law exponent $\kappa$. To extract the evolution of the cut-off scale $\Lambda(t)$ from the lattice data, we exploit the fact that the spectrum shown in Fig. \ref{fig:Spectra} shows a self-similar behavior. To illustrate this we first note that we can parametrize the entire spectrum as
\begin{eqnarray}
f(t,\p)=\alpha^{-c}\left(\frac{\Lambda_s(t)}{|\p|}\right)^\kappa  C(t,|\p|/\Lambda(t))\;,
\end{eqnarray}
where the first part corresponds to the infrared power-law and the second part regulates the ultra-violet behavior such that $C(t,|\p|/\Lambda(t))=1$ for $|\p|<\Lambda(t)$, whereas it drops off quickly for $|\p|>\Lambda(t)$ as observed in Fig. \ref{fig:Spectra}. Here $\kappa\simeq4/3$ is assumed to be constant in time. If we assume further that the shape of the cut-off function is independent of time, i.e.
\begin{eqnarray}
\label{eq:cutoff}
C(t,|\p|/\Lambda(t))\simeq C(|\p|/\Lambda(t))\;,
\end{eqnarray}
and the scales $\Lambda(t)$ and $\Lambda_s(t)$ evolve according to a power law in time, i.e.
\begin{eqnarray}
\label{eq:LScaling}
\Lambda(t)&\sim&Q~(t/t_{\text{Start}})^{\alpha}\;, \\
\label{eq:LSScaling}
\Lambda_s(t)&\sim&Q~(t/t_{\text{Start}})^{\beta}\;,
\end{eqnarray}
we find that the distribution function is self-similar in the sense that for $s>0$ one finds
\begin{eqnarray}
\label{eq:invScaling}
f(st,s^{\alpha}\p)=s^{(\beta-\alpha)\kappa} f(t,\p)\;.
\end{eqnarray}
Here it is crucial to rescale momenta according to $s^\alpha$, whereas the evolution time scales with $s$ in order to reproduce the correct cut-off behavior. Therefore Eq. (\ref{eq:invScaling}) is particularly sensitive to the scaling of the hard scale $\Lambda(t)$ an can be used to extract the scaling exponent $\alpha$.  We note that the assumption (\ref{eq:cutoff}) only concerns the behavior of $p>\Lambda(t)$, while for $p<\Lambda(t)$ this property is satisfied by construction. Hence the condition (\ref{eq:cutoff}) reflects the assumption, that the fall-off above the cut-off scale is universal and we will turn back to this point when we contrast the above analysis with lattice data. Also the scaling behavior of $\Lambda_s(t)$ readily emerges from Fig. \ref{fig:LambdaSoft}, where we confirmed $\beta\kappa=(\kappa-4)/7$ in agreement with the analysis in Sec. \ref{sec:kin}, where we obtained the exponents $\alpha=1/7$ and $\beta=(1-4/\kappa)/7$. \\
\\
In order to investigate to which degree this self-similarity is featured by the lattice data, we invert Eq. (\ref{eq:invScaling}) such that we expect the spectra at different times to collapse on a universal curve. As we are particularly interested in the evolution of the hard scale $\Lambda(t)$, we find it convenient to consider moments of the distribution function
\begin{eqnarray}
f^{(n)}(t,\p)=|\p|^n f(t,\p) 
\end{eqnarray}
which are more sensitive to hard modes than the distribution function itself. Setting $s=t/t_0$ the scaling relation for the moments $f^{(n)}(t,\p)$ then takes the form
\begin{eqnarray}
f_0^{(n)}(\p)=(t/t_0)^{(4-n)/7}f^{(n)}(t,(t/t_0)^{1/7}\p)
\end{eqnarray}
where we used the values of $\alpha,\beta$ from Sec. \ref{sec:kin} and abbreviated $f_0^{(n)}(\p)=f^{(n)}(t_0,\p)$ for arbitrary $t_0>t_{\text{start}}$. The rescaled moments of the lattice data are shown in Fig. \ref{fig:Moments} for $n=2,3$ and $4$ and times $Qt=1000-8000$, where we used $Qt_0=1000$ as the normalization. As the moments are more sensitive to the hard tail of the distribution, we used $N=96$ and $Qa=0.33$ in our simulations to achieve a larger momentum cut-off . One observes that the rescaled data collapses approximately on a single curve. This provides strong evidence that the hard scale $\Lambda(t)$ indeed follows a power law behavior with $\Lambda(t)\sim Q~(Qt)^{1/7}$ as discussed in Sec. \ref{sec:kin}. In particular the position and amplitude of the peak, which are very sensitive to $\alpha=1/7$, coincide for all curves. Above the peak we observe minor deviations from the universal behavior. This can be attributed to a non-universal fall-off of the cutoff function for $p>\Lambda(t)$, however we can also not completely rule out the presence of lattice artifacts in the high momentum region.
\begin{figure}[t]
\includegraphics[width=0.45\textwidth]{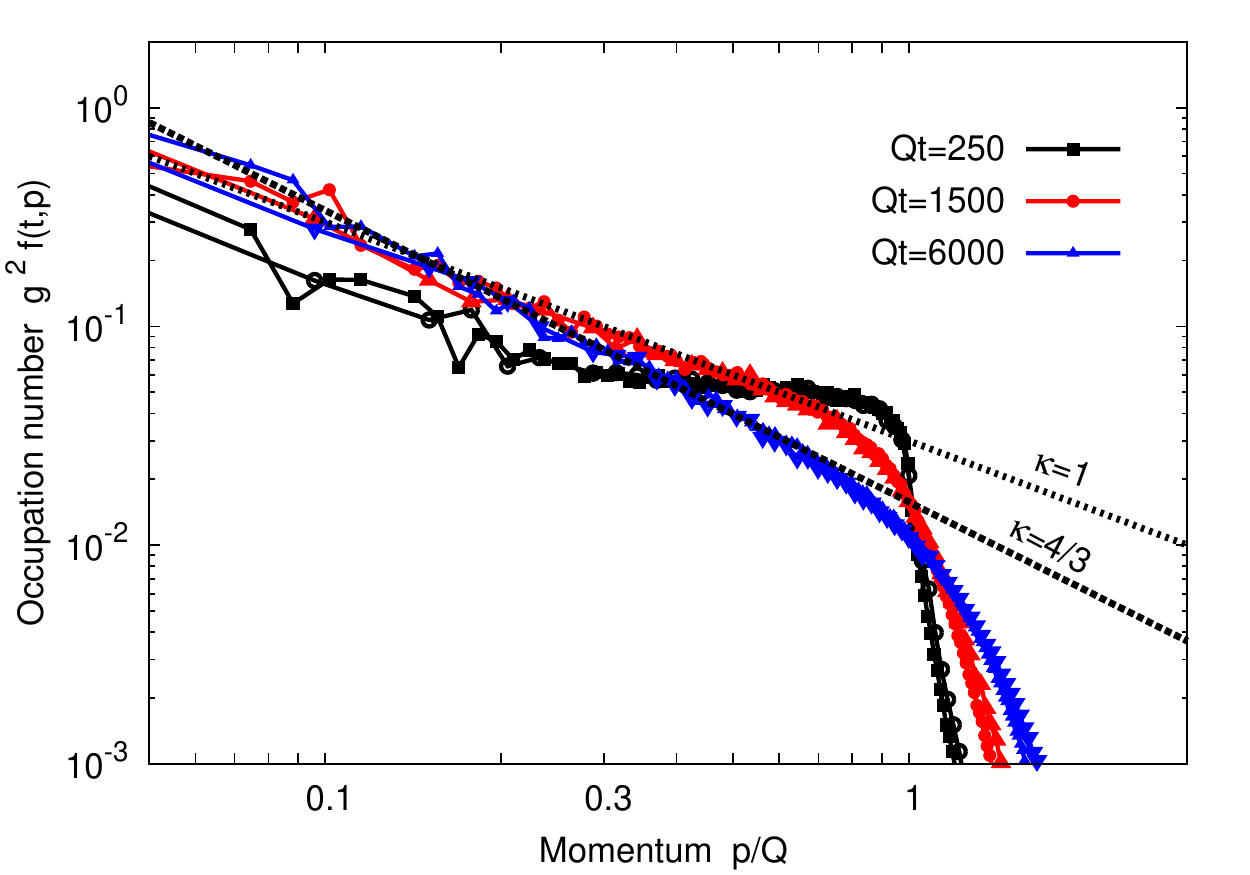}
\caption{\label{fig:SpectraC005} (color online) Spectrum of excitations for $n_0=0.05$ at different times of the evolution. In addition to $N=96$ and $Qa=0.66$ data (open symbols) we also show results with reduced statistics for $N=128$ and $Qa=1$ (filled symbols). One observes that the evolution of modes with $p\simeq Q$ is much slower as compared to the $n_0=1$ case. As a consequence only the soft sector is subject to changes due to interactions at early times, while at later times hard modes are also affected. In the transition region the spectrum shows a thermal shape $\kappa\simeq1$ at some point of the evolution. Nevertheless the evolution at late times proceeds via a turbulent cascade with $\kappa\simeq4/3$. This behavior is indicated by the black dashed lines. }
\end{figure} 
\subsubsection{Overoccupied systems \\ ($f_0(p)\sim\alpha^{-c},~0<c<1$)}
We now extent our previous analysis to systems with initial occupancies which are still parametrically large but now smaller than $\alpha^{-1}$. We will in the following denote the initial occupancy by $n_0=\alpha^{1-c}$ and we study systems with $n_0=0.05,0.1,0.2,0.5$ for a fixed value of the strong coupling constant $\alpha=10^{-6}$ such that the classicality condition $n_0\gg\alpha$ is always satisfied. We will first discuss the case $n_0=0.05$ as an example and then turn to a global analysis of all data. We note that the qualitative behavior on large time scales is very similar for all values of $n_0$ considered here.\\
\\
We proceed as previously and first study the evolution of the spectrum of the correlation functions. In Fig. \ref{fig:SpectraC005} we present snapshots of the spectrum of $f(t,\p)$ as defined in Eq. (\ref{eq:defOcc}) for initial occupation $n_0=0.05$. At early times we observe an increase of occupancies in the soft sector, whereas the occupation for modes with $p\simeq Q$ remains almost unaffected. This is in accordance with our expectations from the kinetic theory analysis (c.f. Sec. \ref{sec:kin}). The evolution at late times proceeds, similarly to the case $n_0=1$, as a turbulent cascade ($\kappa\simeq4/3$) towards the ultraviolet. The transition region between early and late times is characterized by a softening of the spectrum, where the initially flat distribution changes its shape to a power-law distribution with exponent $\kappa\simeq4/3$. Even though the classical thermal value $\kappa\simeq1$ is featured at some time during this transition, the system subsequently continues to evolve towards $\kappa\simeq4/3$ at later times. This provides yet another strong piece of evidence that a non-thermal exponent $\kappa\simeq1$ is indeed favored during the far from-equilibrium evolution.\\
\\
\begin{figure}[t]
\includegraphics[width=0.45\textwidth]{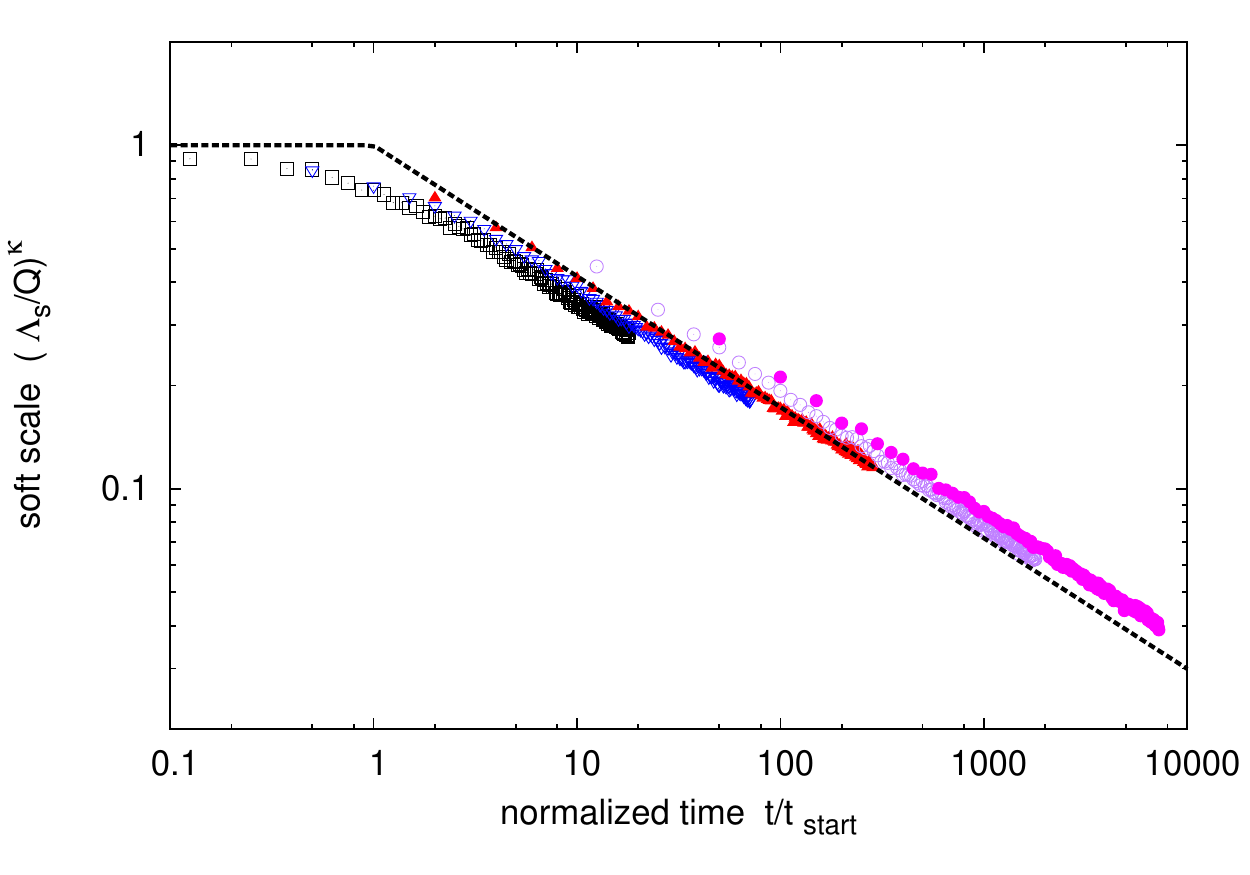}
\caption{\label{fig:tScaling} (color online) Evolution of the soft scale $(\Lambda_s/Q)^\kappa$ normalized by the initial occupation $n_0$ as a function of the normalized time $t/t_{\text{Start}}$ for different initial occupations $n_0=0.05,0.1,0.2,0.5$ and $1$ (bottom to top). The quantity $(\Lambda_s/Q)^\kappa$ is obtained from the occupation of modes with $p\simeq0.9Q$. The black dashed line portraits the result of the kinetic theory analysis (c.f. Sec. \ref{sec:kin}). At late times $t\gg t_{\text{Start}}$ one observes a power law dependence independent of the choice of $n_0$ in the considered range of parameters.}
\end{figure} 
When comparing the results in Fig. \ref{fig:SpectraC005} with the evolution in the case $n_0=1$, we observe that the evolution proceeds significantly slower in the case of reduced occupancy. This is expected given the smaller interacting rates for lower occupancies. The kinetic theory analysis presented in Sec. \ref{sec:kin} here makes a strong statement about the occupancy dependence, based on the parametric dependence of the momentum diffusion parameter $\hat{q}\sim n_0^2Q^3$, which controls the rate of interactions at the hard scale. Accordingly the natural time scale for the system to evolve is therefore no longer set by $Q^{-1}$ but rather by the 'scattering time' $t_{\text{Start}}\sim Q^{-1}n_0^{-2}$ which initially controls the rate of interactions at the hard scale (see also Ref. \cite{AK1}). In order to analyze whether this behavior is displayed by the lattice data, we chose to investigate the time evolution of the soft scale $(\Lambda_s/Q)^\kappa$ for different initial occupancies. This quantity can be obtained from the occupation of modes with $p\simeq0.9~Q$ and is shown in Fig. \ref{fig:tScaling} as a function of time in units of the 'scattering time'. One observes that around $t\simeq t_{\text{Start}}$ the transition from an approximately constant behavior to a power-law decay takes place. The time dependence at late times $t\gg t_{\text{Start}}$ is well described by a power law, where $(\Lambda_s/Q)^\kappa\sim(t/t_{\text{start}})^{(\kappa-4)/7}$ with $\kappa\simeq4/3$ as indicated by the black dashed line in Fig. \ref{fig:tScaling}. Concerning the dependence on the initial occupancy $n_0$, we find that the expected scaling holds approximately as the data shown in Fig. \ref{fig:tScaling} nearly collapses on a single curve.  However one does observe a residual dependence of the overall amplitude of $(\Lambda_s/Q)^\kappa$ on the initial occupancy, which we found to be rather sensitive to the details of the initial conditions.
\section{Summary and Discussion}
\label{sec:conc}
In this paper we studied the evolution of weakly coupled non-abelian plasmas, which are characterized by parametrically large occupancies. In Sec. \ref{sec:kin} we extended the kinetic theory discussion of Refs. \cite{AK1,Blaizot} to account for the presence of a non-thermal fixed point, characterized by a power law distribution with exponent $\kappa$. We found that for $\kappa<3/2$ the dynamics is dominated by hard modes and the analysis in Refs. \cite{AK1,Blaizot} can be extended to non-thermal systems in a straightforward way. In particular the evolution of the hard scale $\Lambda$ and the occupancies at the hard scale $(\Lambda_s/\Lambda)^\kappa$, which are relevant to estimate the thermalization time are unaffected by this generalization. In Sec. \ref{sec:lat} we studied the time evolution of the system within classical-statistical real-time lattice simulations. We presented strong evidence that the evolution proceeds as a turbulent cascade with exponent $\kappa\simeq4/3$, where the system exhibits a self-similar behavior. This is further supported by the results reported in Ref. \cite{BergesTurb}, where $\kappa=4/3$ has been obtained from both numerical and analytical considerations.  When taking this into account in the kinetic description, we found good overall agreement with the lattice data. In summary our results suggest a 'turbulent thermalization' mechanism, which is very similar to observations in early-universe cosmology \cite{cosmo} and systems of cold-atomic gases \cite{qgas,qgas1,qgas2}.\\
\\
In view of the early time dynamics of relativistic heavy-ion collisions it is important to take into account the longitudinal expansion of the system. The latter renders the system anisotropic on large time scales and additional effects such as plasma instabilities are expected to play an important role \cite{AK2}. It is therefore not possible to directly  transfer our results to expanding systems and separate numerical studies will have to be performed. While significant progress has recently been made in simulations of expanding systems in scalar quantum field theories \cite{Scalars}, present simulations in non-abelian gauge theories \cite{Glasma} have not yet been able to resolve this question and we expect future studies to clearify the situation.\\
\\
\textit{Acknowledgement}: The author likes to thank J. Berges, L. McLerran, D. Sexty and in particular R. Venugopalan for insightful discussion and collaboration on related projects. The author also thanks Brookhaven National Lab for hospitality during his stay, where this work was performed. This work was supported in part by BMBF grant 06DA9018 and by HGS-HIRe for FAIR.
\section*{Appendix A: Initial conditions}
The initial conditions for our classical-statistical lattice simulation are chosen to mimic a quasi particle picture, where we consider a superposition of modes $a_{\lambda,a}(t_0,\p)$ labeled by the color index $a=1,...,N_c^2-1$, the polarization $\lambda=1,2$ and spatial momentum $\p$. The mode functions $a_{\lambda,a}(t_0,\p)$ are the ones of the free theory, which satisfy the gauge-condition $\p_ia_{\lambda,a}^i(t_0,\p)=0$ as well as the abelian part of the gauss-law constraint $\p_i \partial_t\left.a_{\lambda}^i(t,\p)\right|_{t=t_0}=0$ individually for each mode. The initial occupation of modes is then determined by the relations
\begin{eqnarray}
\label{eq:latticeic}
\left.\langle a_{\lambda,a}(t,\p) a^*_{\lambda',b}(t',\p)\rangle\right|_{t=t'=t_0}&=&\delta_{ab}\delta_{\lambda\lambda'}N_\p/2\omega_\p \nonumber \\
\left.\partial_t\langle a_{\lambda,a}(t,\p) a^*_{\lambda',b}(t'\p)\rangle\right|_{t=t'=t_0}&=&0 \nonumber \\
\left.\partial_t\partial_{t'}\langle a_{\lambda,a}(t,\p) a^*_{\lambda',b}(t'\p)\rangle\right|_{t=t'=t_0}&=&\delta_{ab}\delta_{\lambda\lambda'}\omega_\p N_\p/2
\nonumber \\
\end{eqnarray}
where $N_\p=f_0(\p)+1/2$ is the initial distribution of excitations. We consider systems with parametrically large occupation according to
\begin{eqnarray}
f_0(\p)=\alpha^{-c}~\theta(Q-|\p|) 
\end{eqnarray}
where $\theta(x)$ is the Heavyside step function, $\alpha\ll1$ is the strong coupling constant and $0<c<1$ as discussed previously. The characteristic mode energy $\omega_\p$ in Eq. (\ref{eq:latticeic}) is initially chosen as
\begin{eqnarray}
\omega_\p=\sqrt{\p^2+\alpha^{1-c}Q^2}
\end{eqnarray}
 where we added a mass term proportional to the Debye screening mass to avoid infrared divergencies. Finally we fix the non-abelian Gauss-law constraint by use of a standard relaxing algorithm, which we apply to the electric field variables.

\end{document}